\documentclass[5p,review]{elsarticle}%
\usepackage{amsmath} % AMS Math Package
\usepackage{amsthm} % Theorem Formatting
\usepackage{amssymb}	% Math symbols such as \mathbb
\usepackage{graphicx} % Allows for eps images
\usepackage[usenames,dvipsnames]{pstricks}
\usepackage{pst-grad} % For gradients
\usepackage{pst-plot} % For axes
\usepackage{color}
\usepackage{wasysym}
\usepackage{url}
\usepackage{bbold}
\usepackage{natbib}
\biboptions{sort&compress}
\usepackage{bm}
\newcommand{\markup}[1]{{\color{black}{#1}}}
\renewcommand{\v}[1]{\ensuremath{\mathbf{#1}}} % for vectors
\newcommand{\gv}[1]{\ensuremath{\mbox{\boldmath$ #1 $}}} % for vectors of Greek letters
\newcommand{\avg}[1]{\left< #1 \right>} % for average
\renewcommand{\d}[2]{\frac{d #1}{d #2}} % for derivatives
\newcommand{\pd}[2]{\frac{\partial #1}{\partial #2}} % for partial derivatives
\newcommand{\pdd}[2]{\frac{\partial^2 #1}{\partial #2^2}} % for double partial derivatives
\newcommand{\ket}[1]{\left| #1 \right>} % for Dirac bras
\newcommand{\bra}[1]{\left< #1 \right|} % for Dirac kets
\renewcommand{\div}[1]{\gv{\nabla} \cdot #1} % for divergence
\let\baraccent=\= % rename builtin command \= to \baraccent
\renewcommand{\=}[1]{\stackrel{#1}{=}} % for putting numbers above =

\theoremstyle{definition}

\theoremstyle{remark}


\begin{document}

\title{Quantum Confinement in Nonadditive Space with a Spatially Dependent Effective Mass for Si and Ge Quantum Wells}

\author[immcnr]{E. G. Barbagiovanni \corref{cor1}}
\ead{eric.barbagiovanni@ct.infn.it}

\author[fort]{R. N. Costa Filho}

\cortext[cor1]{Corresponding author}

\address[immcnr]{MATIS IMM-CNR, via S. Sofia 64, 95123 Catania, Italy}

\address[fort]{Departamento de F\'{i}sica, Universidade Federal do Cear\'{a}, Caixa Postal 6030, Campus do Pici, 60455-760 Fortaleza, Cear\'{a}, Brazil}

\date{\today}

\begin{abstract}

We calculate the effect of a spatially dependent effective mass (SPDEM) [adapted from R. N. Costa Filho et al. Phys. Rev. A., \textbf{84} 050102 (2011)] on an electron and hole confined in a quantum well (QW). In the work of Costa Filho et al., the translation operator is modified to include an inverse character length scale, $\gamma$, which defines the SPDEM. The introduction of $\gamma$ means translations are no longer additive. In nonadditive space, we choose a `skewed' Gaussian confinement potential defined by the replacement $x\rightarrow\gamma^{-1}\ln(1+\gamma x)$ in the usual Gaussian potential. Within the parabolic approximation $\gamma$ is inversely related to the QW thickness and we obtain analytic solutions to our confinement Hamiltonian. Our calculation yields a reduced dispersion relation for the gap energy ($E_G$) as a function of QW thickness, $D$: $E_G\sim D^{-1}$, compared to the effective mass approximation: $E_G\sim D^{-2}$. Additionally, nonadditive space contracts the position space metric thus increasing the occupied momentum space and reducing the effective mass, in agreement the relation: $m_o^{*-1}\propto\pdd{E}{\v{k}}$. \markup{The change in the effective mass is shown to be a function of the confinement potential via a point canonical transformation.} Our calculation agrees with experimental measurements of $E_G$ for Si and Ge QWs.

\end{abstract}


\maketitle

\section{Introduction\label{intro}}

The introduction of a spatially dependent effective mass (SPDEM) in the Hamiltonian has benefited research from quantum gravity to condensed matter \cite{CostaFilho:2011, Mazharimousavi:2012}. In semiconductor physics, a SPDEM can be applied to doped semiconductors, or when there exists a graded potential \cite{vonRoos:1983, Young:1989, Geller:1993}. The von Roos Hamiltonian \cite{vonRoos:1983} expresses a general form of the kinetic energy operator with a scalar SPDEM:
\begin{equation}\label{vonRH}
\begin{array}{ll}
&T=-\frac{\hbar^2}{4}\left(m(\v{r})^{\alpha}\div m(\v{r})^{\beta} \div m(\v{r})^{\delta}+\right.\\
&\left. + m(\v{r})^{\delta}\div m(\v{r})^{\beta} \div m(\v{r})^{\alpha} \right);
\end{array}
\end{equation}
with the constraint $\alpha+\beta+\delta=-1$. Apart from the constraint, the parameters $\alpha$, $\beta$, and $\delta$ are arbitrary. Therefore, the parameters of Eq. \eqref{vonRH} suffer from an ordering ambiguity, which has lead to a debate in the literature \cite{vonRoos:1983, Cruz:2007, Mustafa:2009, Mustafa:2006, Mustafa:2013}. In the work of Ref. \cite{Mustafa:2013}, the author claims to clear up the ordering ambiguity for the case of a harmonic potential. 

\markup{A point canonical transformation (PCT) of the von Roos Hamiltonian generates a constant mass and an effective potential \cite{CostaFilho:2013, vonRoos:1983}. Simarly, }permutations in the crystal potential of an intrinsic semiconductor influence the charge carrier's (electrons and holes) effective mass (EM) \cite{JohnPeter:2008, Khordad:2010}. Likewise, at the interface of a nanostructure (NS) where the crystal potential will change, possibly abruptly, the EM will change. The Bastard-type boundary conditions (B.Cs) are appropriate when there is a change in the EM at the interface of a NS \cite{Bastard:1981, Chetouani:1995, Ganguly:2006, Moskalenko:2007}:
\begin{equation}\label{BastBCs}
\frac{1}{m_A}\d{F^A}{z}=\frac{1}{m_B}\d{F^A}{z};
\end{equation}
for the envelope function, $F$, and mass, $m$, in material $A$ or $B$ denoted along the $z$-direction. Eq. \eqref{BastBCs} is a generalization of the usual continuous B.Cs ($F^A=F^B$ and $\d{F^A}{z}=\d{F^A}{z}$) and can be derived from Eq. \eqref{vonRH} with the constraint: $\alpha=\beta=0$, and $\delta=-1$ \cite{Bastard:1981}. The Bastard type B.Cs have been directly applied to the problem of quantum confinement (QC) in NSs \cite{Moskalenko:2007}.
 
Proper treatment of EM in a NS is an unresolved problem. The EM ($m_o^{*}$) is important for theoretical models \cite{Tomic:2011, Delerue:2004, Niquet:2000_1} as it is related to the hopping parameter and carrier mobility in the tight-binding model \cite{Seino:2012}. There is experimental \cite{Barbagiovanni:2012, Seas:1997} and theoretical \cite{Seino:2011} evidence that the EM should depend on NS dimension. Fundamentally, since QC increases the occupied momentum space \cite{Barbagiovanni:2012}, one does expect an increase in $m_o^{*-1}(\propto\pdd{E}{\v{k}})$. Furthermore, a change in $m_o^*$ modifies the Bohr radius, therefore, altering the regime in which QC effects can be observed. The challenge is that experimental measurements of $m_o^*$ in a NS are model dependent \cite{Lockwood:1996, Rossner:2003} and it is difficult to theoretically scale the EM to low dimensions \cite{Barbagiovanni:2012, Seino:2012}. However, the EM provides a natural framework to incorporate the influence of a modified crystal potential due to the interface, which is not adequately accounted for in theoretical models \cite{Delerue:2004, Barbagiovanni:2013, Seino:2011}.

Theoretical work that explicitly considers a spatial dependence in the EM studied how to represent the effective Hamiltonian \cite{Mazharimousavi:2012, Mustafa:2013}, or the problem of donor impurities \cite{JohnPeter:2008, Khordad:2010}. In a few reports, a dimensionally dependent EM was deduced from a fit with experimental data \cite{Quang:2008} or by including higher order corrections \cite{Nehari:2008}. Density functional theory (DFT) was used to calculate the EM from the density of states \cite{Seino:2011}. However, it is not clear how to model the interface nor the excited states within DFT \cite{Barbagiovanni:2011, Moskalenko:2007}.

In this report, we present a new approach to the problem of the low dimensional EM and the interfacial energy, which is important for carrier transport applications in NSs \cite{Titova:2011}. We consider a SPDEM in the confinement Hamiltonian, derive the ground state envelope functions, and calculate the gap energy ($E_G$) as a function of quantum well (QW) thickness. Our formalism does not suffer from an ordering ambiguity, because it was derived from first-principles. The SPDEM was derived by considering a particle confined in nonadditive space. Our confinement potential was chosen based on the properties of nonadditive space and lead to analytic solutions for the confinement Hamiltonian. \markup{In this formalism, we demonstrate the equivalence between the confinement potential (which is related to the interfacial energy) and EM via a PCT.}
 
\section{Theory\label{theory}}

\subsection{Theoretical Background\label{theorybkg}}
The formalism presented in this manuscript was adapted from Ref. \cite{CostaFilho:2011}, which is summarised here. Costa Filho et al. introduced a characteristic inverse length scale, $\gamma$, in the translation operator defined by \cite{CostaFilho:2011}:
\begin{equation}\label{Transl}
\mathcal{T}_{\gamma}(a)\ket{x}=\ket{x+a+\gamma a x}.
\end{equation}
$\gamma$ mixes the displacement, $a$, of a carrier particle with the original position, $x$. Two successive infinitesimal translations was given by \cite{CostaFilho:2011}:
\begin{equation}\label{TranslTwo}
\mathcal{T}_{\gamma}(dx')\mathcal{T}_{\gamma}(dx'')=\mathcal{T}_{\gamma}(dx'+dx''+\gamma dx'dx'');
\end{equation} 
which demonstrates that translations are no longer additive. We denote this nonadditive space as `$\gamma$-space.' From Eq. \eqref{Transl} the momentum operator was derived \cite{Sakurai:1994}:
\begin{equation}\label{pGamma}
\hat{p}_{\gamma}=-i\hbar(1+\gamma x)\d{}{x}.
\end{equation}
In the limit $\gamma \rightarrow$ 0, the standard momentum operator was recovered. The condition that $\hat{p}_{\gamma}$ remains hermitian was maintained with a modification to the completeness relation:
\begin{equation}\label{GammaNorm}
\mathbb{1}=\int \frac{dx}{1+\gamma x} \ket{x}\bra{x};
\end{equation} 
where $\mathbb{1}$ is the unit matrix. 

\markup{In accordance with Eq. \eqref{pGamma}, the kinetic operator was modified to read:
\begin{equation}\label{KineGamma}
\hat{T}=\frac{\hat{p}_{\gamma}^2}{2m_o^*}=-\frac{\hbar^2}{2m(x)}\pdd{}{x}-\frac{\hbar^2}{2}\pd{}{x}\left(\frac{1}{2m(x)}\right)\pd{}{x}
\end{equation}
where $m_o^*$ is the bulk EM and a} SPDEM was identified as:
\begin{equation}\label{SPDEM}
m(x)=\frac{m_o^*}{\left(1+\gamma x\right)^2}.
\end{equation}
Note that while Eq. \eqref{KineGamma} is not of the form given by Eq. \eqref{vonRH}, it was derived from first-principles. Depending on the value of $\gamma$, position space is either contracted or dilated, and was shown to affect the energy spectrum and probability amplitude of a free-particle and an infinitely confined particle, respectively \cite{CostaFilho:2011}. \markup{Eq. \eqref{Transl} states that translations in $\gamma$-space are coupled with the origin. This coupling modifies the momentum (Eq. \eqref{pGamma}) of the charge carriers, which is equivalent to a change in the effective mass (Eq. \eqref{SPDEM}). In Sec. \ref{defgamma}, we will derive a functional form for $\gamma$ and explain the physical significance. First, we define our confinement Hamiltonian.} 

\subsection{Quantum Confinement in $\gamma$-Space\label{QC}}

\markup{In this section, we define the} Hamiltonian for an electron and hole confined by a `skewed' Gaussian potential:
\begin{equation}\label{confHamil}
\begin{array}{ll}
&\mathcal{H}=\int d^3x \psi^{\dagger}(x)\left(\hat{T}\right)\psi(x)-\\
&-V_o\int d^3x \psi^{\dagger}(x)\exp\left(\frac{1}{2R^2\gamma^2}\ln^2(1+\gamma x)\right)\psi(x);
\end{array}
\end{equation}
where $V_o$ is the depth of the confinement potential, \markup{$\psi(x)$ is a field operator,} and $R$ is the radius of the QW. $V_o$ is defined as the energy difference between the QW and the matrix material at the conduction band minimum (CBM) or the valence band maximum (VBM) for an electron ($V_{o,e}$) or hole ($V_{o,h}$), respectively. The confinement potential in Eq. \eqref{confHamil} is simply a Gaussian potential:
\begin{equation}\label{Gauss}
V_G=-V_o\exp\left(-x^2/2R^2\right);
\end{equation}
where $x\rightarrow\gamma^{-1}\ln(1+\gamma x)$ \markup{introducing a `skew' in the Gaussian potential. This modification to the confinement potential is chosen for two reasons: 1) it is equivalent to the PCT discussed in Ref. \cite{CostaFilho:2013}, and 2) it leads to a definition for $\gamma$ and thus an analytic solution for the confinement energy.} For our present purposes we do not consider exciton effects, which constitute a small correction to the energy spectrum in Si and Ge NSs \cite{Barbagiovanni:2013, Barbagiovanni:2012}.

Eq. \eqref{confHamil} is solved in the two-band effective mass approximation (EMA). The EMA is well suited to model the Si/SiO$_2$ interface \cite{Gusev:2013} through the use of the envelope function approximation (EFA). Furthermore, the two-band EMA agrees well with experimental results compared to a multi-band approach \cite{Moskalenko:2007, Barbagiovanni:2012, Nguyen:2012}. The details of this method can be found in Refs. \cite{Moskalenko:2007, Gusev:2013, Barbagiovanni:2011, Barbagiovanni:2013}. 

Note that, all functions considered here are properly normalized according to a complete set of states given by Eq. \eqref{GammaNorm}. Additionally, we assume the Bloch functions carry the normalization condition:
\begin{equation}\label{Bloch}
\int_{\Omega} dx\; u^*_{k,i}(x)u_{k,j}(x)=\Omega\delta_{i,j}, 
\end{equation}
where the integration is over the unit cell volume, $\Omega$. This assumption is justified in the EFA, because the Bloch functions vary over the length of the lattice spacing.

The confinement potential in Eq. \eqref{confHamil} can be approximated within the parabolic approximation \cite{Adamowski:2000} as:
\begin{equation}\label{parobVc}
\begin{array}{ll}
V_C&=-V_o\exp\left(\frac{1}{2R^2\gamma^2}\ln^2(1+\gamma x)\right)\\
&\approx -V_o + \frac{V_o}{2\gamma^2R^2}\ln^2(1+\gamma x)\\
&=-V_o + \frac{1}{2}\frac{m_o^*\omega^2}{\gamma^2}\ln^2(1+\gamma x);
\end{array}
\end{equation}
where:
\begin{equation}\label{omega}
\omega^2=\frac{V_o}{m_o^*R^2}.
\end{equation}
Eq. \eqref{parobVc} is similar to the potential of a harmonic oscillator, where $\omega$ represents the oscillator frequency. The constant $V_o$ term in Eq. \eqref{parobVc} is absorbed into the total energy. With the confinement potential in the form of Eq. \eqref{parobVc}, solutions to Eq. \eqref{confHamil} are straightforward. We derived the normalized envelope function given by:
\begin{equation}\label{psi}
\Psi_{i}(x)=\left(\frac{2}{\sigma_i^2\pi}\right)^{1/4}\exp\left(-\frac{1}{2\sigma_i^2\gamma_i^2}\ln^2(1+\gamma_i x_i)\right);
\end{equation}
where $\sigma^2=\frac{\hbar}{m_o^*\omega}$ is the Gaussian width parameter\markup{, and the subscript $i$ is contained in either the conduction or valence band. To proceed further in the formalism we need a definition for $\gamma$. This step is carried out in Sec. \ref{defgamma}, and the energy eigenvalues are discussed in Sec. \ref{QCresults}, but first we mention some qualitative features of the formalism.}

\begin{figure}[h]
\includegraphics[scale=.7]{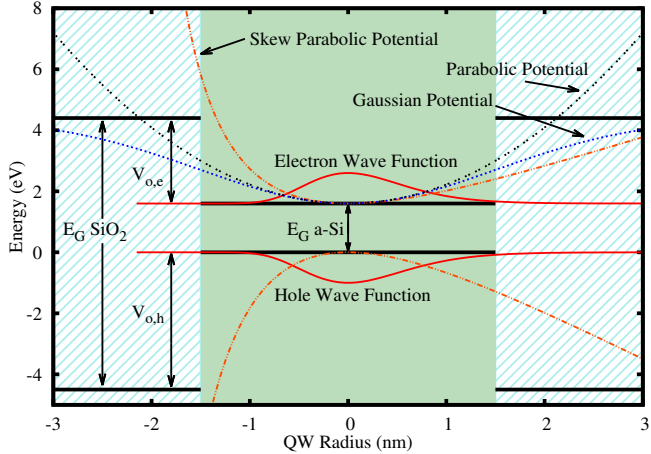}
\caption{Energy diagram for an electron and hole in a 3 nm a-Si QW (solid green region between -1.5 to 1.5 nm) confined by SiO$_2$ (stripped blue region). Horizontal black lines indicate the CBM and VBM energy level, with $E_G$ a-Si=1.6 eV, $E_G$ SiO$_2$=8.9 eV. $V_{o,e}$=2.8 eV and $V_{o,h}$=4.5 eV are the band offsets between a-Si and SiO$_2$ for the electron and hole, respectively. All functions are plotted with a band offset given by the a-Si/SiO$_2$ interface. The electron and hole wave-functions are given by Eq. \eqref{psi} ($y$-axis for $\Psi$ is not depicted). The Gaussian potential is given by Eq. \eqref{Gauss}, and the parabolic potential is approximated from Eq. \eqref{Gauss}. The skew parabolic potential is given by Eq. \eqref{parobVc}. \label{Gaussband}}
\end{figure}
The energy diagram for an electron in the conduction band (CB) and a hole in the valence band (VB) is plotted in Fig. \ref{Gaussband} for a 3 nm amorphous Si (a-Si) QW confined by SiO$_2$. $V_{o,e}$=2.8 eV and $V_{o,h}$=4.5eV are calculated according to the electron affinity rule, which agrees with experimental results \cite{Seguini:2011}. These values assume an abrupt interface, which is not always true at the Si/SiO$_2$ interface \cite{Lockwood:1999, deSousa:2002}. We will discuss this assumption further in Sec. \ref{QCresults}. Fig. \ref{Gaussband} shows the Gaussian potential (Eq. \eqref{Gauss}), along with its parabolic approximation, and our `skew parabolic potential' (Eq. \eqref{parobVc}). Around $x$=0 the three potentials agree. As the particle moves toward the QW interface in the positive direction the skew parabolic potential is lower than the other two potentials, while the opposite is true in the negative direction. This condition is represented in the asymmetric spread of the wave function across the QW thickness for both the electron and hole. The asymmetry is a reflection of the half interval ($\gamma^{-1},\infty$) on which the particle is bounded from Eq. \eqref{SPDEM}.

\subsection{Definition of $\gamma$\label{defgamma}}

\markup{Here we find a formalism for $\gamma$ and discuss the physical meaning of $\gamma$-space. Solutions for $\gamma$ are not straight forward. We consider first an analogous formalism provided by Costa Filho et al.\cite{CostaFilho:2013}. In their work, the authors considered a harmonic oscillator in $\gamma$-space, with the Hamiltonian:
\begin{equation}\label{GammaSHO}
\mathcal{H}_{\gamma}^{SHO}=\frac{\hat{p}_{\gamma}^2}{2m_{o}^*}+\frac{1}{2}m_{o}^*\omega^2x^2.
\end{equation}
From Eq. \eqref{GammaNorm} we see that the integration measure is not one-to-one between $\gamma$-space and Cartesian-space, i.e. the Jacobian is not equal to one. Therefore, in accordance with Eq. \eqref{GammaNorm}, a modified coordinate system is defined in Ref. \cite{CostaFilho:2013} as: 
\begin{equation}\label{eta}
\eta(x)=\gamma^{-1}\ln(1+\gamma x). 
\end{equation}
From Eq. \eqref{eta}, the `canonical coordinate' is written as: $x(\eta)=(\exp(\gamma\eta)-1)\gamma^{-1}$. A PCT is now performed on Eq. \eqref{GammaSHO}, which yields the Hamiltonian in $\eta$-space:
\begin{equation}\label{GammaSHOeta}
\mathcal{H}_{\eta}^{SHO}=\frac{-\hbar^2}{2m_{o}^*}\pdd{}{\eta}+V_{eff}(\eta).
\end{equation}
$V_{eff}(\eta)$ is derived from the SHO potential in Eq. \eqref{GammaSHO} and is given by:
\begin{equation}\label{moorseeta}
V_{eff}(\eta)=\frac{m_{o}^*\omega^2}{2\gamma^2}(1-e^{\gamma\eta})^2.
\end{equation}
With Eq. \eqref{GammaSHOeta} we have a} Morse oscillator in $\eta$-space for a  particle with constant mass, in perfect analogy with Ref. \cite{vonRoos:1983}. Therefore, a Harmonic oscillator in $\gamma$-space with a SPDEM is equivalent to a Morse oscillator in $\eta$-space with a constant mass. 

\markup{Before discussing the formalism for $\gamma$, it is important to review some basic features of the PCT. In the above formalism, we performed a canonical transformation within position, $x$, space on $\mathcal{H}_{\gamma}^{SHO}$ (Eq. \eqref{GammaSHO}). The effect of the transformation is to remove the spatial dependence from the EM and produce an effective potential in $\eta$-space (Eqs. \eqref{GammaSHOeta} and \eqref{moorseeta}). Therefore, this transformation `mimics' the relationship between the crystal potential and the EM. That is, within the $\v{k}\cdot\v{p}$ formalism the energy due to the periodic crystal potential is modelled by replacing the free electron mass with an EM \cite{Barbagiovanni:2013}. Hence, Eq. \eqref{moorseeta} in $\eta$-space incorporates the spatial dependence from the EM in Eq. \eqref{GammaSHO}.

In this work, we wish to address the question of how the EM is modified in low-dimension. Specifically, we want to address how the interfacial confinement potential modifies the EM, in analogy with the crystal potential and the EM. Therefore, to address this point we use the same PCT from Ref. \cite{CostaFilho:2013} in our confinement potential, which is our motivation for Eq. \eqref{confHamil}. 

Now, we may employ the formalism of Ref. \cite{CostaFilho:2013} for use here.} We have demonstrated that our skewed-harmonic potential (Eq. \eqref{parobVc}) is related to the Morse potential through a PCT. The Morse potential describes bonding in a diatomic molecule and has the form:
\begin{equation}\label{Morse}
V_M(r) = V_o(1-e^{-\alpha r})^2;
\end{equation}
where $r$ is the distance between the atoms and $\alpha$ is an inverse length parameter. If we expand Eq. \eqref{Morse} and Eq. \eqref{parobVc} around the origin, they agree up to second order, because both potentials describe a particle bounded on a half interval. \markup{By comparing Eq. \eqref{moorseeta} with Eq. \eqref{Morse}, we immediatly identify \cite{CostaFilho:2013}:
\begin{equation}\label{gam1}
\gamma^2=\frac{m_o^*\omega^2}{2V_o}.
\end{equation}}
Using our expression for $\omega$ (Eq. \eqref{omega}), we obtain a simple expression for $\gamma$:
\begin{equation}\label{gam}
\gamma^2=\frac{1}{2R^2}
\end{equation}
Notice that this definition is in agreement with Eq. \eqref{Gauss}.

\begin{figure}[h]
\includegraphics[scale=.85]{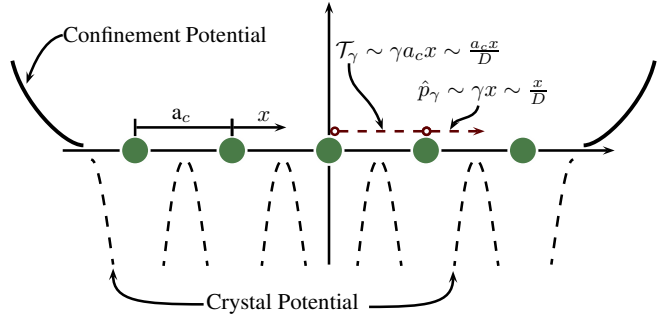}
\caption{Cartoon depicting the effect of $\gamma$ on translations, $\mathcal{T}_{\gamma}$, and the momentum operator, $\hat{p}_{\gamma}$. Atoms are separated by the lattice spacing, $a_c$. Charge carriers experience a crystal potential and a confinement potential, $V_C$. For $\mathcal{T}_{\gamma}$ on the order of the lattice spacing, depicted by the dashed red line, there is an increase in $\hat{p}_{\gamma}$ as $D$ decreases.\label{gammaspace}}
\end{figure}
\markup{Finally, we address the physical meaning of $\gamma$-space. First, we notice that $\eta(x)$ (Eq. \eqref{eta}) increases sub-linearly with $x$. Since $\gamma$ is inversely related to the thickness of the QW (Eq. \eqref{gam}), this sub-linear dependence is enhanced as the thickness of the QW is reduced. This behaviour is the source of the `skew' in $V_C$ seen in Fig. \ref{Gaussband}. Therefore, the influence of the interface potential is enhanced as the QW thickness is reduced. Next, if recall Eqs. \eqref{Transl}, \eqref{pGamma}, and \eqref{SPDEM}, we can understand how the energy spectrum and the EM are modified as a result of $\gamma$. In Fig. \ref{gammaspace}, we have depicted a cartoon of $\mathcal{T}_{\gamma}$ (Eq. \eqref{Transl}), and $\hat{p}_{\gamma}$ (Eq. \eqref{pGamma}) with respect to the crystal lattice. The atoms in this cartoon are separated by the lattice spacing, $a_c$, and the charge carriers are displaced by $x$. The crystal potential is the background energy of the charge carriers that gives rise to the EM. The interface energy resulting from $V_C$ is related to the value of $\gamma$. As $\gamma$ increases (or the QW thickness is reduced) translations on the order of $a_c$ are increasingly coupled to the origin. This coupling is given by the factor $\gamma a_c x$ represented by the dashed red line in Fig. \ref{gammaspace}. Similarly, the momentum of the charge carriers is modified due to $\gamma$. The modified momentum is a function of the SPDEM from Eq. \eqref{SPDEM}, and thus is related to $V_C$, analogous the effect of the crystal potential.}

\section{Results\label{results}}

\markup{First, we briefly examine the effect of $\gamma$ on $V_C$ and the probability amplitude with respect to the interface. Second, we will discuss the change in the EM, which is related to $V_C$, see Sec. \ref{defgamma}. Finally, we will study the results for the confinement energy.}

\subsection{Confinment Potential in $\gamma$-Space \label{contr}}

\begin{figure}[h]
\includegraphics[scale=.7]{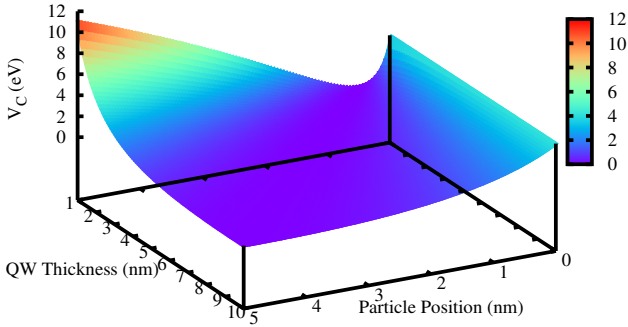}
\caption{Confinement potential Eq. \eqref{parobVc} as a function of QW thickness and particle position. \label{PotentialQWX}}
\end{figure}

\begin{figure}[h]
\includegraphics[scale=.7]{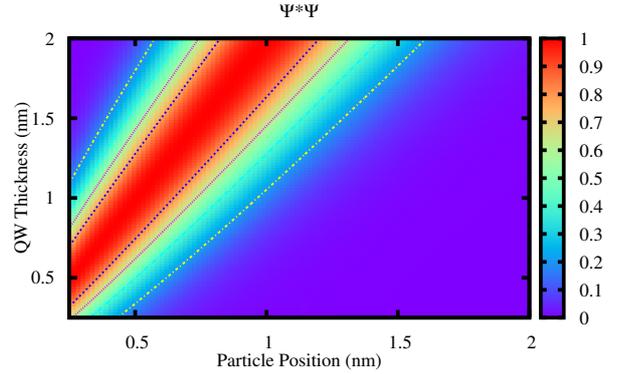}
\caption{Probability amplitude of $\Psi$ Eq. \eqref{psi} as a function of QW thickness and particle postion. \label{ProbabilityQWX}}
\end{figure}

The dimensionally dependent confinement potential (Eq. \eqref{parobVc} with Eq. \eqref{gam}) is plotted in Fig. \ref{PotentialQWX} as a function of QW thickness and particle position. As the dimension of the QW increases, the curvature of $V_C$ decreases thus decreasing the confinement strength. For $D\rightarrow \infty$, $V_C$ goes to zero. The behaviour of $\Psi$ follows the same trend as $V_C$. Fig. \ref{ProbabilityQWX} plots the probability amplitude of $\Psi$ as a function of QW thickness and particle position. The width of the probability amplitude increases with QW thickness and is skewed in the positive direction.

\begin{figure}[h]
\includegraphics[scale=.7]{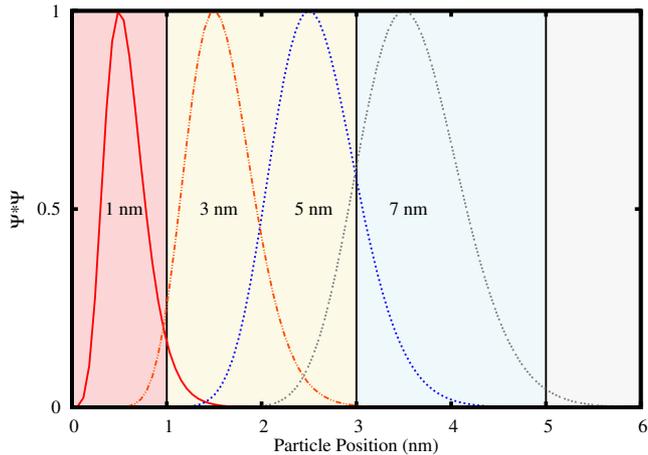}
\caption{Probability amplitude for an electron in different QW thicknesses confined by the a-Si/SiO$_2$ interface parameters: $V_{o,e}$=2.8 eV and $V_{o,h}$=4.5eV. The probability amplitudes are centred within the QW. The thickness of each QW is indicated in the plot under the respective curve. The vertical lines at 1, 3, and 5 nm are to guide the eye to the interface of the different QW thicknesses. \label{ProbabilityXD}}
\end{figure}

\markup{Now, recall Sec. \ref{defgamma}, in $\gamma$-space the weight of our Jacobian lead to a sub-linear increase in $\eta(x)$.} This effect manifests in the full width half max of $\Psi$, which increases sub-linearly with QW thickness. A closer examination of the probability amplitude with QW thickness reveals how the confined particles are spread within the QW. Fig. \ref{ProbabilityXD} plots the probability amplitude for four different QW thicknesses: 1, 3, 5, and 7 nm. For a 1 nm thick QW, Fig. \ref{ProbabilityXD} demonstrates that the tail of the probability amplitude extends into the SiO$_2$ interface. While, for 5 and 7 nm the probability amplitude does not extend into the matrix material. The width of $\Psi$ is inversely related to the QW thickness according to the definition of $\gamma$. Hence, there is a reduced tunnelling probability as the QW thickness increases. It has been demonstrated \cite{Gusev:2013} that the effect carrier tunnelling into the oxide matrix is small, in agreement with our results. \markup{Therefore, as the QW thickness is reduced and $V_C$ increases, the carrier coupling with the interface increases, which modifies the EM as we will discuss in the next section.}

\subsection{Gap Energy in $\gamma$-Space\label{QCresults}}

\begin{figure}[h]
\includegraphics[scale=.7]{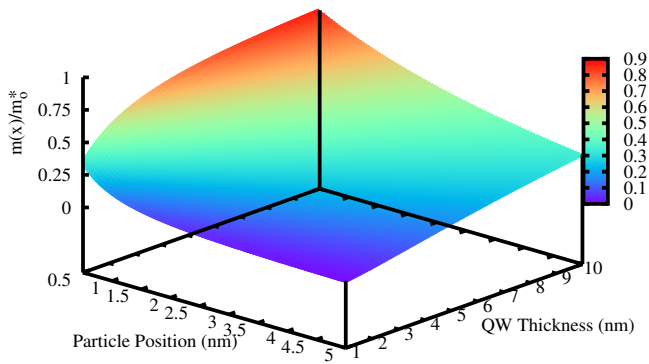}
\caption{Plot of Eq. \eqref{SPDEM} as a function QW thickness and particle postion. \label{EM}}
\end{figure}

The confined carrier's EM is modified in $\gamma$-space. The SPDEM is plotted in Fig. \ref{EM} as a function of particle position and QW thickness. Similar to the asymmetry in $\Psi$, there is an asymmetry in the SPDEM. As $x\rightarrow 0$, the bulk EM is recovered. In the limit that the QW thickness goes to infinity, the bulk effective mass is recovered, because $\gamma\rightarrow 0$. On the other hand, as the QW thickness is reduced the confined particle `feels' a stronger confinement potential thus lowering the EM. \markup{This result is analogous to the effect of the crystal potential on the free electron mass, see Sec. \ref{defgamma}.} These results agree with the prediction: $m_o^{*-1}\propto\pdd{E}{\v{k}}$, which states that a strongly de-localized particle in momentum space (i.e. a strongly confined particle) exhibits a reduced EM.

We can understand the reduced EM in more detail by considering $\Psi$ in momentum space. The Fourier transform in $\gamma$-space is given by:
\begin{equation}\label{fourier}
F(k)=\int \frac{1}{1+\gamma x} \Psi(x) \exp\left(-i\frac{k}{\gamma}\ln(1+\gamma x)\right) dx.
\end{equation}
This definition yields:
\begin{equation}\label{fk}
F(k)\sim \sqrt[4]{R}\exp\left(-Rk^2/2\right).
\end{equation}
Eq. \eqref{fk} demonstrates a reduced spread in momentum space with increasing QW thickness, as expected. Furthermore, compared to a particle in a Gaussian well (with $\gamma = 0$: $F_{\gamma=0}(k)\sim \sqrt{R}\exp\left(-R^2k^2/2\right)$), the spread in momentum space is increased. This result is a direct consequence of contracted $\gamma$-space from Eq. \eqref{GammaNorm} compared to ordinary position space. Therefore, we observe an increasing spatial confinement of the carrier particles, thus increasing the occupied momentum space, as the QW thickness is reduced. \markup{Hence, the EM is lowered in $\gamma$-space and varies with QW thickness, which modifies the dispersion of the carrier particles and increases the confinement energy.}

\markup{We solve Eq. \eqref{confHamil} in the electron-hole basis, $\Phi$, defined above the ground state, $\Phi_0$ \cite{Barbagiovanni:2011}. The eigenvalues are given by: $E_G(D)=E_G(\infty)+ \bra{\Phi}\mathcal{H}\ket{\Phi}$, which gives:
\begin{equation}\label{QCEnergy}
E_G(D)=E_G(\infty)+\frac{\hbar}{\sqrt{2}D}\left[\sqrt{\frac{V_{o,e}}{m_{o,e}^*}}+\sqrt{\frac{V_{o,h}}{m_{o,h}^*}}\right];
\end{equation}
where $E_G(\infty)$ is the bulk $E_G$, and we have written the variation in $E_G$ in terms of the QW thickness, $D$. We use the EM values of Ref. \cite{Barbagiovanni:2012}. Note that this formalism is easily extended into higher dimensions, because the wave-function is separable. If we assume the same form of $V_C$ for each confinement dimension, in 2D (quantum wires) and in 3D confinement (quantum dots), Eq. \eqref{QCEnergy} is multiplied by two and three, respectively.}

\begin{figure}[h]
\includegraphics[scale=.7]{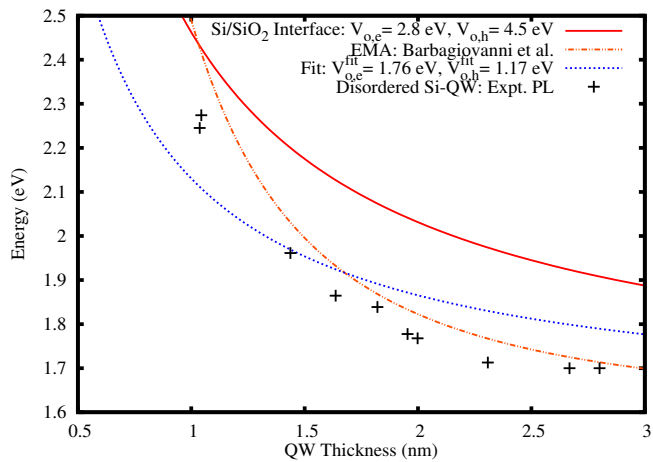}
\caption{Plot of the theoretical calculation Eq. \eqref{QCEnergy} with the experimental data from Ref. \cite{Lu:1995}. Bulk $E_G$ of disordered Si is 1.6 eV. Theoretical curves use the a-Si/SiO$_2$ interface parameters and fitted interface parameters as labelled in figure. Infinite confinement EMA model is shown for comparison from Barbagiovanni et al. \cite{Barbagiovanni:2012}.\label{QCGammafit}}
\end{figure}

\begin{figure}[h]
\includegraphics[scale=.7]{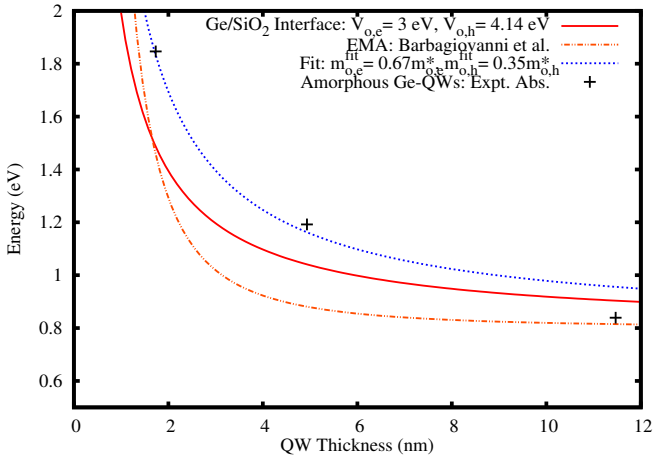}
\caption{Plot of the theoretical calculation Eq. \eqref{QCEnergy} with experimental data from Ref. \cite{Cosentino:2013}. Bulk $E_G$ of amorphous Ge is 0.8 eV. Theoretical curves use the a-Ge/SiO$_2$ interface parameters and fitted mass parameters as labelled in figure. Infinite confinement EMA model is shown for comparison from Barbagiovanni et al. \cite{Barbagiovanni:2012}.\label{QCGammafitGe}}
\end{figure}

We plotted Eq. \eqref{QCEnergy} alongside a similar EMA calculation using an infinite confinement potential \cite{Barbagiovanni:2012} in Figs. \ref{QCGammafit} and \ref{QCGammafitGe}. Our calculation is compared with experimental data for high quality Si \cite{Lu:1995} and Ge \cite{Cosentino:2013} QWs. An important feature of our model is the reduced dispersion of the confined particles: $E_G\sim D^{-1}$ in Eq. \eqref{QCEnergy} versus $E_G\sim D^{-2}$ in the EMA \cite{Barbagiovanni:2012}. The cause of the reduced dispersion is discussed above. 

Fig. \ref{QCGammafit} shows experimental data for a disordered Si-QW from Ref. \cite{Lu:1995}. The QWs were prepared using molecular beam epitaxy. Our calculation overestimates $E_G$ when the a-Si/SiO$_2$ interface parameters are used. It is important to note that the QW interface is composed all Si sub-oxide states \cite{Lu:1996}. Sub-oxide states reduce the interface energy offset, $V_o$. To account for the change in the interface energy we fitted Eq. \eqref{QCEnergy} to the experimental data using fitting parameters with $V_{o,e}$ and $V_{o,h}$. From the fit we obtain: $V^{fit}_{o,e}$ = 1.76 eV and $V^{fit}_{o,h}$ = 1.17 eV (Fig. \ref{QCGammafit}). The fitting parameters give an interface $E_G$ of 4.53 eV, which is reduced from the SiO$_2$ $E_G$ at 8.9 eV. The interface energy is in excellent agreement with experimental measurements, which report a 3 to 5 eV reduction from the SiO$_2$ interface \cite{Muller:1999, Yamashita:2006}. Furthermore, there is an increased dispersion in the experimental data at small QW dimensions, which is missed in our model. This feature could be a result of an increased concentration of oxygen vacancy defect states at the interface \cite{Barbagiovanni:2013, Lockwood:2004_1}, due to an error in the measurement of the QW dimension \cite{Lu:1996, Garrido:2004, Barbagiovanni:2012}, or possibly because the carriers experience a stronger confinement energy from an additional mechanism not considered here. Nonetheless, our model is in good agreement with the experimental data. 

Experimental data for a-Ge QWs prepared by magnetron sputtering with an SiO$_2$ interface \cite{Cosentino:2013} is shown alongside our theoretical prediction in Fig. \ref{QCGammafitGe}. In their work, Cosentino et al. prepared high quality Ge QWs with a sharp interface by not annealing the QWs after deposition (see transmission electron microscopy image in Ref. \cite{Cosentino:2013}). Their unique fabrication method reduced the formation of sub-oxide states at the interface. Therefore, we can assume that the interface energy is approximately given by the gap offset between a-Ge and SiO$_2$, indicated in Fig. \ref{QCGammafitGe}. Our calculation is in good agreement with the experimental data using the a-Ge/SiO$_2$ interface parameters. The SPDEM model is an improvement over the infinite confinement model shown in Fig. \ref{QCGammafitGe}, but still underestimates the experimental $E_G$. 

To account for the difference in $E_G$ Cosentino et al. fitted our infinite confinement EMA to their experimental data. This fit models a reduction in the bulk reduced EM, $\mu^*_{o}$, given by: $\mu^{fit}_{o}=0.45 \mu^*_{o}$. If we perform the same fit using our $\gamma$ model, we obtain: $m^{fit}_{o,e}=0.67m^*_{o,e}$ and $m^{fit}_{o,h}=0.35m^*_{o,h}$, which gives $\mu^{fit}_{o}=0.23 \mu^*_{o}$. This value is in good agreement with the report of Ref. \cite{Cosentino:2013}. Discrepancies arise because of the different behaviour of holes versus electrons in a NS due to the interface \cite{Seguini:2013}. As NS dimension is reduced holes become more localized than electrons \cite{Seino:2012} due to pinning with interface states \cite{Seguini:2013, Saar:2005, Garrido:2000, Barbagiovanni:2012}. This behaviour is not accounted for in our present model. Additionally, Cosentino et al. pointed out that Ge NSs experience stronger confinement energies compared to similar Si NSs \cite{Cosentino:2013}. The reason for this difference is due to the larger Bohr radius in Ge compared to Si \cite{Barbagiovanni:2012, Barbagiovanni:2013}. Therefore, a Ge NS experiences a larger reduction in the EM with NS dimension compared to Si, which explains the need to perform a fit using the Ge EM.

Our SPDEM results agree with other experimental reports. In the work of Ref. \cite{Rossner:2003}, the authors reported a reduced electron EM, by fitting temperature dependent Shubnikov-de Haas oscillations for Ge QWs, from the bulk value ($m^*_{o,e}$=1.08$m_o$ and $m^*_{o,e}$=0.56$m_o$ in Si and Ge, respectively, where $m_o$ is the free electron mass) at 0.08$m_o\approx$ 0.14$m^*_{o,e}$. The tunnelling EM was reported to be 0.09$m_o$ in amorphous Si \cite{Shannon:1993}. Temperature dependent photoluminescence measurements placed the electron EM at 0.014$m_o$ for Ge/Si superlattices \cite{Yang:2004}. The reported values are smaller than our fitted values above, because these experiments involved Ge QWs confined by Si versus the case of SiO$_2$ considered above. A Si matrix implies reduced interface energy from SiO$_2$. 

\section{Discussion\label{disc}}

Our results demonstrate a reduced dispersion relation compared to the standard EMA and an increased overall confinement energy in $\gamma$-space. The choice for a dimensionally dependent $V_C$ (Eq. \eqref{parobVc}) reduces the confinement strength as the QW dimension increases. At the interface ($x=0$) $m(0)= m_o^*$. For a small value of $\gamma$ (i.e. a large QW), the change in the SPDEM is small as the carriers oscillate within the QW (Fig. \ref{EM}). On the other hand, a large value of $\gamma$ implies a large reduction of the EM, due to the large spread in momentum space Eq. \eqref{fk}.

The features of this model are unique in that we are able to model the effects of a modified $V_C$ and EM. However, there are some important corrections we are not concerned with in the present report. As discussed above and seen in Fig. \ref{PotentialQWX}, the interface is not considered to vary symmetrically across the well. The reason for this phenomena is due to the definition in Eq. \eqref{Transl}. Recall, that the confined particles are bound on the half interval $\left(-\gamma ^{-1},\infty\right)$. Other functional forms for a SPDEM have been reported in the literature \cite{Cruz:2007}. However, to the best of our knowledge only Eq. \eqref{SPDEM} can be derived from first principles. 

In the work of Ref. \cite{Cruz:2007}, the authors consider three different forms of the SPDEM, given by:
\begin{subequations}\label{massfunc}
\begin{align}
&m_1(x)=\frac{m^*_o}{1+(\lambda x)^2}\qquad (\textrm{no singularities})\label{no}\\
&m_2(x)=\frac{m^*_o}{(1+\lambda x)^2}\qquad (\textrm{1 singularity})\label{one}\\
&m_3(x)=\frac{m^*_o}{(1-(\lambda x)^2)^2}\qquad (\textrm{2 singularities});\label{two}
\end{align}
\end{subequations}
where $\lambda$ is an arbitrary inverse length parameter. Eq. \eqref{no} represents a free-particle. Our formalism is equivalent Eq. \eqref{one}. Eq. \eqref{two} is bounded on the interval $(-\lambda^{-1}, \lambda^{-1})$ and was used in the work of Ref. \cite{Mustafa:2009} with a von Roos Hamiltonian. A PCT of Eq. \eqref{two} with spatial coordinates given by: $q(x)=\lambda\tanh^{-1}(x/\lambda)$ yields a modified P\"{o}schl-Teller type potential \cite{Mustafa:2009}, see Sec. \ref{defgamma}. The modified P\"{o}schl-Teller potential represents an infinitely confined particle, which does not represent the correct confinement condition for Si and Ge NSs. 

Cruz et al. solved the Hamiltonian:
\begin{equation}\label{CruzH}
H=-\frac{1}{2}m^a\d{}{x}m^{2b}\d{}{x}m^a+V(x);
\end{equation}
where $a+b=-1/2$, for each of the three mass functions in Eq. \eqref{massfunc} using a set of carefully constructed creation-annihilation operators. The normalization condition does not agree with the one used in this work. For the case of $m_2(x)$, Cruz et al. used the same functional form of the confinement potential as in Eq. \eqref{parobVc}, but the units do not agree with ours. For $m_3(x)$, Cruz et al. used a similar P\"{o}schl-Teller type potential as in Ref. \cite{Mustafa:2009}. These formalisms while novel do not yield complete solutions, because $\lambda$ is an arbitrary parameter. There is no restriction that a SPDEM must obey the von Roos Hamiltonian. 

We can qualitatively discuss our results with respect to calculations of the EM with QW thickness, i.e. no spatial dependence. If we consider a fixed position in Eq. \eqref{SPDEM}, say $x\rightarrow \avg{x}$, the average position, then the EM decreases with QW thickness (Fig. \ref{EM}). This behaviour is in agreement with the results of Ref. \cite{Seino:2011} for the electron EM, however, their calculation yields a value larger than the bulk EM. A decreasing EM with QW thickness agrees well with experimental results \cite{Barbagiovanni:2012}, see Secs. \ref{intro}, and \ref{QCresults}. However, the calculation of Ref. \cite{Seino:2011} does not qualitatively agree with the experimentally observed reduced EM from the bulk value, where our model does. On the other hand, the hole is known to become more localized with respect to the electron \cite{Seino:2011}, due to pinning with the interface \cite{Barbagiovanni:2012}. Therefore, we would expect an increase in the hole EM as NS dimension is reduced in agreement with \cite{Seino:2011}. In Refs. \cite{Nehari:2008, Quang:2008}, a dimensionally dependent EM was derived within the EMA by fitting with the tight binding model. Those results do not agree with the results of this work and with Ref. \cite{Seino:2011}.

Finally, it is important to comment on corrections that can be made to the present model. We discussed above the interface asymmetry. Furthermore, in this report, we assumed an isotropic CB EM, because the experiments considered here used amorphous QWs. Using the anisotropic values (m$_{\parallel}$ and m$_{\perp}$) for the electron EM can be included in this model. This correction is small \cite{Moskalenko:2007, Barbagiovanni:2012} and does not change our general conclusions. As commented above and in Sec. \ref{QCresults}, the pinning effect of the hole is not currently accounted for. This correction should be treated with caution since the behaviour of the hole depends on the structure of the interface \cite{Seguini:2013, Saar:2005}. The interface structure will change with QW thickness \cite{Barbagiovanni:2013}. Here, the interfacial energy is averaged over the QW thickness through the $V_o$ parameters. Such an effect is a challenge for current theoretical models, but was considered in the work of Refs. \cite{Sousa:2000, deSousa:2002}. Further corrections to the model may include a Bastard type B.C derived from Eq. \eqref{KineGamma}.

\section{Conclusions\label{conc}}

We studied the effect of a SPDEM in a two-band EM QC model. The SPDEM was derived by introducing an inverse characteristic length in the translation operator thus defining nonadditive space ($\gamma$-space) for the confined particles. The confinement energy was derived in $\gamma$-space using a `skewed' parabolic potential, $V_C$. The novel choice in $V_C$ yielded analytic solutions for the confinement Hamiltonian and treated the interface energy. The results demonstrate a reduced dispersion relation for the confined particles compared to an infinite confinement model and an increase in the confinement energy. The increased confinement energy was understood as a result of the reduced EM, due to the change in $V_C$. The EM of the confined particles was modified as a result of a contracted space due to $\gamma$. In $\gamma$-space, the confined particles occupy more momentum space as compared to a Gaussian envelope with $\gamma\rightarrow$ 0. This increased spread in momentum space was the cause of the reduced EM. Our results demonstrated an improvement to the EMA model for confined particles. These results are a promising first step toward the development of a better model for the EM parameter in low-dimensional theories. Work is being carried out to test these results within the $\v{k}\cdot\v{p}$ model. Eq. \eqref{pGamma} may also be included in tight-binding, or pseudo-potential methods. Additional work is being completed to test the dimensional dependence ($D^{-2}$ versus $D^{-1}$) of our model versus the EMA and tight binding model for $E_G$ across a range of NS dimensions.  

\section*{Acknowledgements}
R.N.C F acknowledges funding from CNPq and the Brazilian agency Funda\c{c}\~{a}o Cearense de Apoio ao Desenvolvimento Cient\'{i}fico e Tecnol\'{o}gico (FUNCAP) through PRONEX (Grant No. PR2-0054-00022.01.00/11).

\section*{References}
%\bibliography{refr}

\end{document}